\documentclass{article}

\usepackage[preprint]{neurips_2023}

\usepackage[utf8]{inputenc} % allow utf-8 input
\usepackage[T1]{fontenc}    % use 8-bit T1 fonts
\usepackage{hyperref}       % hyperlinks
\usepackage{url}            % simple URL typesetting
\usepackage{booktabs}       % professional-quality tables
\usepackage{amsfonts}       % blackboard math symbols
\usepackage{nicefrac}       % compact symbols for 1/2, etc.
\usepackage{microtype}      % microtypography

\usepackage{amssymb}
\usepackage{amsmath}
\usepackage{amsfonts}
\usepackage{graphicx}
\usepackage{xcolor}
\usepackage{fancyhdr}
\newtheorem{definition}{Definition}

\begin{document}

\title{ Supply Chain Coordination Mechanism Design  \\  \large Consensus Planning Protocol Meets Vickrey-Clarke-Groves Mechanism  }

\author{Dirk Bergemann \thanks{Email: \texttt{dirk.bergemann@yale.edu}}  \\
  \And
  Garrett van Ryzin \thanks{Email: \texttt{garrett.vanryzin@gmail.com}} \\
  \And
 Jiaxuan Li \thanks{Email: \texttt{lijiaxuan0529@gmail.com}}
 }

%\date{5/15/2026}
\maketitle

\maketitle

\begin{abstract}
  This paper introduces the theoretical framework for combining Vickrey-Clarke-Groves (VCG) mechanisms with the Consensus Planning Protocol (CPP) to enable truthful and efficient collaboration between a retailer and vendors to lower joint Cost to Serve (CtS). We demonstrate how this integration preserves both dominant-strategy incentive compatibility and efficiency in high-dimensional environments. We further introduce an activity fee design to improve its revenue property for the retailer while maintaining the mechanism's desirable properties. This CPP-VCG framework serves as the theoretical foundation for designing collaborative mechanisms to coordinates distributed, agent-based optimization between retailers and suppliers. 
  
\end{abstract}

\section{Introduction}
This paper presents the theoretical foundation powering a next-generation collaboration technology that lowers cost to serve (CtS) for both Vendors and a retailer by using distributed, agent-based optimization. The technology addresses a critical challenge in supply chain coordination where retailers and suppliers independently pursue local optimizations, commonly resulting in sub-optimal outcomes for both parties. The Consensus Planning Protocol (CPP) (see \cite{cppguide}, \cite{cpp2022}) - a distributed, agent-based optimization framework - has demonstrated success in internal coordination. 
In this paper, we extend the CPP framework to enable coordination between a retailer agent and external agents (e.g., Vendor agents). However, to apply CPP with external parties, we need to overcome two critical problems: 1) ensuring each party has an incentive to submit truthful data to their agents and 2) ensuring each party has the incentive to participate in the coordination mechanism. We solve these problems by combining CPP with the Vickrey-Clarke-Groves (VCG) mechanism in a design that achieves both incentive compatibility and efficiency. The resulting VCG transfer payment —ensures truthful participation and redistributes coordination gains between parties.

We first develop the connection between CPP and VCG and the implications for vendor coordination. We then propose to modify the gross utility of a retailer by an additive factor to improve its revenue property (utility boosting). 
This additive boosting is particularly attractive, resulting in a simple modified VCG mechanism in which supplier's pay an ``activity fee" in addition to their usual VCG payment. The mechanism
will still be incentive compatible, meaning suppliers have no advantage in misreporting their costs and benefits. We then show how to extend the mechanism from static to dynamic environments, as demand predictions and supply conditions can change over time. Lastly, we show that a "menu of contracts" can serve as an alternative design for lower-dimensional environments.

\section{Model}
\subsection{Payoff Environment}
We begin with a stylized setting to anchor our discussions. Suppose a retailer buys from a supplier under its usual just-in-time (JIT) policy but would like to find a more efficient way to trade. The JIT policy described here is a stylized model for analytical tractability. Actual buyer-supplier interactions involve additional complexities not captured in this framework. For simplicity, assume the supplier provides a single product (ASIN) to a retailer and decisions are made on a weekly basis, where $t$ indexes weeks. To lower total cost, both parties seek to collaborate on a supply plan over a planning horizon of $T$ weeks. We assume there are $I$ potential inbound nodes, indexed by $i=1,\ldots,I$. Then the supply plan is defined by an $I \times T$ matrix:
\[
x=[x_{it}],
\]
where $x_{it}$ is the number of units of product delivered to inbound node $i$ in week $t$. 
The utility of a retailer for a plan $x$ is given by
\[
u_A(x, \theta_A) = r_A(x,\theta_A) - c_A(x, \theta_A),
\]
where $r_A(x,\theta_A)$ is the unconstrained net revenue a retailer receives on sales of the product over the planning horizon; $c_A(x, \theta_A)$
is the total supply chain costs it incurs as a result of the plan $x$, which includes costs related to lost-sales/backlogged demand, inventory holding, in-bounding and out-bounding, and $\theta_A$ denotes a retailer's private information on its costs and benefits. 

Similarly, the utility of the supplier $S$ for a plan $x$ is given by
\[
u_S(x, \theta_S) = r_S(x,\theta) - c_S(x, \theta_S),
\]
where $r_S(x,\theta_S)$ is the net revenue the supplier receives on deliveries of their product to a retailer over the planning horizon; $c_S(x, \theta_S)$
is the supply chain costs the supplier incurs as a result of the plan $x$, which includes cost related to production, inventory holding, shipment processing and transporting to inbound locations, and $\theta_S$ denotes the supplier's private information on its costs and benefits. 

The retailer and the supplier would each like to increase their individual utility by collaborating on a joint plan, and ideally maximize the benefits from such collaboration. The potential gains from collaboration are maximized by the socially efficient plan:
\[
x^*(\theta) = \arg \max_x \{u_A(x, \theta_A) + u_S(x, \theta_S) \} ,
\]
where the socially efficient plan depends on the vector of private information $\theta = (\theta_A,\theta_S)$.
We explore next how such an outcome can be achieved.

\subsection{CPP}

In the basic CPP setting, there is a public variable vector $x$ which affects the utility of multiple agents, indexed by $m=1,\dots,M$. Agents must reach consensus on the value of $x$. Agent $m$ utility for $x$ is given by solving:
\[
u_m(x, \theta_m) = \max_{y_m \in Y_m} u_m(x,y_m,\theta_m) ,
\]
\noindent where $y_m$ is a vector of private variables of agent $m$, $Y_m$ is a set of constraints on the private variables of agent $m$, and $u_m(x,y_m)$ is agent $m$’s utility, which is a joint function of both the public and private variable choices. All these components of the agent’s utility maximization problem are private information of the agent, which is denoted $\theta_m$. Let $u$ denote the vector of the agents utility functions.

The CPP problem then solves the following social welfare maximization problem:
\[
 \max_x \sum_{m=1}^M u_m(x,\theta_m) .
\]
This problem is solved iteratively using CPP as described in \cite{cppguide}. In each iteration, a central coordinator queries the agents and ask them to return their preferred value (``best response"):
\[
\tilde{x}_m = \arg \max_x \left\{ u_m(x,\theta_m) - \pi_m^T x + \| x - z\|^2_2 \right\} ,
\]
where $\pi_m$ is a vector of public variable prices for agent $m$ and $z$ is a proposed consensus value for $x$. The prices and proposed consensus values are updated by the coordinator at each iteration using an alternating direction method of multipliers (ADMM), see~\cite{boyd2011distributed}.

For concave utilities, the iterations provably converges to an optimal solution.
Let $x^*(u)$ denote an optimal solution to this problem, where the notation makes explicit that the optimal solution is a function of the vector of agent utility functions.

\subsection{Mechanism Design}
In mechanism design, it is known that $\left( i\right) $\ social efficiency, 
$\left( ii\right) $\ incentive compatibility, $\left( iii\right) $\
participation constraint and $\left( iv\right) $\ budget balance cannot beattained simultaneously, \cite{mysa83}. But if either one of these four
conditions is abandoned, then the other three can be maintained. More
interestingly, we can ask for an optimal solution with respect to one
criterion, while maintaining the other three exactly. This is typically
referred to as a second-best mechanism.

CPP applied to the internal planning can satisfy $\left( i\right) $ social
efficiency, $\left( ii\right) $ incentive compatibility and $\left(
iv\right) $ budget balance. It does not have to worry about the participation
constraint because agents are internal organizations within a retailer who are forced to participate. However, if we deploy CPP to do collaborative planning with external suppliers, then we may insist on $%
\left( ii\right) $\ incentive compatibility, $\left( iii\right) $\
participation constraint and $\left( iv\right) $\ budget balance, and strive to achieve
social efficiency or private efficiency. Often the
later two goals will be close. Besides the relevance of the second-best solution in the presence of an
impossibility result, there is a second important change when moving from
internal to external coordination. Now one agent also
becomes a principal by running the mechanism. This is potentially limiting the
credibility of the principal and the mechanism in general, see \cite{akli20} .

\subsection{VCG Mechanism}
The Vickrey Clarke Groves (VCG) mechanism  achieves social efficiency, incentive compatibility, the participation constraint - though not budget balance. The issue of budget balance will be addressed later. Following the same notation as above for CPP, VCG works as follows:

\begin{definition}[VCG mechanism]   The direct VCG mechanism proceeds in three steps:
    \begin{enumerate}
    \item Each agent $m$ reports its utility function $u_m(x, \theta_m)$. (This reported utility function is the agent’s “bid”.)
    \item The mechanism then computes the socially efficient choice, $x^*(u)$, based on the reported utility functions, $u$, submitted by the agents. 
    \item Agent $m$ is then required to pay the principal a transfer, $t_m$, equal to its externality on other agents
    \[
    t_m = \max_x \sum_{i \neq m}u_i(x,\theta_i) - \sum_{i \neq m}u_i(x^*(u),\theta_i)  .
    \]
    \end{enumerate}
\end{definition}

Again, this mechanism has the following desirable properties: Truthful bidding is a dominant strategy, so every agent has an incentive to report its true utility function regardless of the bidding strategy of other agents (incentive compatibility); $u_m(x^*) - t_m \geq 0$, so agents receive non-negative utility from participating (participation constraint); and total welfare is maximized under the dominant strategy equilibrium (social efficiency).
However, it is not in general budget balanced.Note the close connection to CPP. In Step 1., submitting a utility function $u_m(x,\theta_m)$ is equivalent to submitting a best-response agent to CPP. In Step 2., solving for the socially efficient choice $x^*(u)$ is then simply solving the CPP problem. Lastly, in Step 3, computing the transfer payment for agent $m$ requires solving another CPP problem, with agent $m$ removed.

\section{The Bilateral Trade Case}

We next apply VCG to our initial setting of collaboration between two agents, a retailer and the supplier, denoted $m=A,S$.
The VCG\ mechanism is compensating every participating agent according to
their marginal contribution.

\subsection{VCG Allocations and Payments}
The socially optimal solution is
\begin{equation}
x^{\ast }\left( \theta \right) \in \underset{x\in X}{\arg \max }\left\{
u_{A}\left( x,\theta _{A}\right) +u_{S}\left( x,\theta _{S}\right) \right\} .
\label{eff}
\end{equation}%
Each agent has utility from participating in the VCG mechanism of
\begin{equation*}
u_{A}\left( x,\theta _{A}\right) - t_{A}\left( x,\theta \right)
\end{equation*}%
and 
\begin{equation*}
u_{S}\left( x,\theta _{S}\right)  - t_{S}\left( x,\theta \right) ,
\end{equation*}%
where $t_{m}\left( x,\theta \right) $ are the VCG monetary transfers (from agents to the principal) used to align
preferences. Note in the absence of an agreement (and transfers) there is privately optimal
action that the remaining agent pursues, which in the general case is:
\begin{equation}
x_{-i}^{\ast }\left( \theta _{-i}\right) \in \underset{x\in X}{\arg \max }%
\left\{ \sum_{j\neq i}u_{j}\left( x,\theta _{j}\right) \right\} .
\label{out}
\end{equation}%
In the classical mechanism design setting, the choice of $x$ is controlled by the principal, who is assumed to have all property rights over the items being auctioned and hence can control the outcome independent of which agents participate.
In a consensus planning setting, however, the situation is different; the vector $x$ is a shared plan, and, absent coordination, agents may have different decision rights over how this plan is constructed. For example, in the case of a retailer coordinating with a supplier, the default contract can be a retailer has the right to issue just-in-time (JIT) purchase orders (POs) based on when, where and how many units it wants from a supplier, while the supplier has the right to determine how much of any given PO they are willing/able to fulfill. The composition of these two decisions then determines the actual number of units received by a retailer at any given place and time. More precisely, let $x$ be the supply plan defined in our canonical example. The default arrangement is that a retailer places a matrix of POs $q=[q_{it}]$, where $q_{it}$ is the purchase quantity ordered into node $i$ in week $t$. After receiving the POs, the supplier then responds with a matrix of confirmation decisions $z=[z_{it}]$, where $z_{it} \in [0,1]$ is the fraction of the order fulfilled (the confirmation rate). The final supply plan is then $x = z \odot q$, where ```$\odot$" denotes element-wise multiplication. In contrast, a consensus plan would select the supply plan, $x$, directly.

We can
easily generalize VCG to this case by defining a participation-dependent set in (\ref{out}): $X_{-i}\subseteq X$, and in particular the set $X_{-i}$ can be a singleton $X_{-i}=\left\{ x_{-i}\right\}$ ,
which would reflect a status quo plan. This
highlights the complementary aspect of the relationship, as the absence of a
single agent reduces the choice set of all other agents.

In the current setting we consider two agents, and thus the utility in the absence of one agent is obtained by solving: 
\begin{equation*}
x_{A}\left( \theta _{A}\right) =\ \underset{x\in X_{-S}}{\arg \max } ~ u_{A}\left( x,\theta _{A}\right) ,\ \ \ \ 
x_{S}\left( \theta _{S}\right) =\ \underset{x\in X_{-A}}{\arg \max } ~ u_{S}\left( x,\theta _{S}\right) .
\end{equation*}%
The VCG transfers are then given by 
\begin{equation}
t_{A}\left( x,\theta \right) = u_{S}\left( x_{S}\left( \theta _{S}\right) ,\theta
_{S}\right) - u_{S}\left( x^{\ast }\left( \theta \right)
,\theta _{S}\right) \label{ta}
\end{equation}%
and 
\begin{equation}
t_{S}\left( x,\theta \right) =u_{A}\left( x_{A}\left( \theta _{A}\right) ,\theta
_{A}\right) - u_{A}\left( x^{\ast }\left( \theta \right)
,\theta _{A}\right)  \label{ts}
\end{equation}%
such that each agent internalizes the social welfare, hence the different
subscript on each side of the inequality in (\ref{ta}) and (\ref{ts}).

\subsection{Budget Balance Results}

We can ask when will there be a budget deficit and when will there be a
budget surplus.

\bigskip

\noindent \textbf{Theorem 0 (Budget Balance). }There is a budget deficit
(a retailer pays out more than zero) if and only if:%
\begin{eqnarray}
t_{A}\left( x,\theta \right) +t_{S}\left( x,\theta \right)  &\geq &0  \notag
\\
&\Leftrightarrow &  \label{tx} \\
u_{S}\left( x^{\ast }\left( \theta \right) ,\theta _{S}\right) +u_{A}\left(
x^{\ast }\left( \theta \right) ,\theta _{A}\right)  &\geq &u_{S}\left(
x_{S}\left( \theta _{S}\right) ,\theta _{V}\right) +u_{A}\left( x_{A}\left(
\theta _{A}\right) ,\theta _{A}\right)   \notag
\end{eqnarray}

\bigskip

Thus, when the value of cooperation is larger than the stand alone value. Or
to put it differently when the agents are complements rather than
substitutes. In the trade environment that we are considering we will thus
have typically deficit. By contrast in the auction environment where we have
competition among the agents, we will have a surplus.

\subsection{Benefit of Principal and Agent}

A unique feature of our setting is that a retailer is both a participating agent and the principal.
If we add the benefit of the principal and
the a retailer agent from joining in the mechanism, we obtain
\[
\left( u_{A}\left( x^{\ast }\left( \theta \right) ,\theta _{A}\right)
- t_{A}\left( x^{\ast }\left( \theta \right) ,\theta \right) \right)
+t_{A}\left( x^{\ast }\left( \theta \right) ,\theta \right) +t_{S}\left(
x^{\ast }\left( \theta \right) ,\theta \right) \\
=u_{A}\left( x_{A}\left( \theta _{A}\right) ,\theta _{A}\right) ,
\]
which is just the stand-alone value of a retailer's utility. Thus the VCG\ mechanism is
benefit-neutral for a retailer as whole (principal plus a retailer agent benefit), which implies all the gains from coordination are allocated to the supplier. To confirm this, note
the supplier's transfer to a retailer is
\[
t_{S}\left( x,\theta \right) =u_{A}\left( x_{A}\left( \theta _{A}\right) ,\theta
_{A}\right) - u_{A}\left( x^{\ast }\left( \theta \right)
,\theta _{A}\right) ,
\]
so the supplier's gain in utility from VCG is $ u_{S}\left( x^{\ast }\left( \theta \right), \theta_A \right)  - u_{S}\left( x_{S}\left( \theta _{S}\right) ,\theta_{S}\right) - t_{S}\left( x,\theta \right)  $, which, substituting the above equation, is equal to 
\[
g(\theta) = u_{S}\left( x^{\ast }\left( \theta \right), \theta_S \right) + u_{A}\left( x^{\ast }\left( \theta \right), \theta_A \right) - u_{S}\left( x_{S}\left( \theta _{S}\right) ,\theta_{S}\right) - u_{A}\left( x_{A}\left( \theta _{A}\right) ,\theta_{A}\right) ,
\]
where $g(\theta)$ denotes the total gain from coordination.
If our primary purpose is reducing the cost we impose on suppliers while simultaneously not increasing our own costs, then the classical VCG outcome with a retailer as principal achieves this outcome and is a potentially desirable design. 
However, we may want to tilt the outcome such that benefits are shared between the retailer and the supplier. 
In the next section, we show how to modify the mechanism to enable benefit sharing.

\section{Activity Fee}
Retailers may increase revenue through an activity fee. Market designers can either implement a transaction fee for each individual transaction (e.g., per purchase order) or design a participation fee for multiple transactions over a given time horizon. The ideal approach involves identifying the appropriate level of activity fee that does not disrupt the supplier participation, as it would not introduce any distortion in the allocation decision. However, retailers cannot anticipate the magnitude of gains suppliers derive from coordination mechanisms. Suppliers’ expectations of gains may diverge from the retailer’s forecasts, leading to misalignment. Consequently, pre-negotiated gain-sharing agreements are often impractical.  Instead, empirical studies suggest iterative experimentation—testing fee structures across segments over time—to refine outcomes. For example: adjusting the share for a purchase order, or modifying revenue shares for batched transactions over a specific horizon (e.g., a quarterly period). As the time horizon or the number of products covered increases, uncertainty also increases regarding the activity fee required to ensure efficient supplier participation. If the fee is excessively high, suppliers may decline participation. Furthermore, discrepancies between suppliers’ expectations of coordination gains and retailers’ projections can reduce overall engagement.

\subsection{Sharing Coordination Surplus}

The \emph{first way} to share coordination surplus is to attempt to maintain the efficient policy
(allocation) and only modify the transfers (pay the supplier less). This will
impact the decision by the supplier to participate as an decrease in the
payment has to violate with some frequency of their participation constraint.
Since a priori, we do not know the size of the social gains or how the
social gains are distributed across supplier and a retailer, this method proceeds
with little information, and thus seems fraught.The \emph{second way} is to change the reported values for policies by
a retailer into the system. This will impact the compensation of the supplier, but
also impact the efficiency of the decision. More precisely, we can \emph{%
understate} the benefit for planned buying and/or \emph{overstate} the
benefits form just in time ordering. Either way, we therefore bias the
decision towards just in time ordering, but whenever we depart from just in
time ordering we are guarantee to receive a share of the the gains. 

\subsection{Bias for Just in Time}

If we return to the determination of the transfers in (\ref{ta}) and (\ref%
{ts}) we see that the supplier's transfer to a retailer is %
\begin{equation}
t_{S}\left( x,\theta \right) = u_{A}\left( x_{A}\left( \theta _{A}\right) ,\theta
_{A}\right) - u_{A}\left( x^{\ast }\left( \theta \right)
,\theta _{A}\right) ,  \label{ts1}
\end{equation}%
then we can either understate $\left( \underline{u}\right) $\ the gains from
the coordinated buying: 
\begin{equation*}
u_{A}\left( x^{\ast }\left( \theta \right) ,\theta _{A}\right) >\underline{u}%
_{A}\left( x^{\ast }\left( \theta \right) ,\theta _{A}\right) ,
\end{equation*}%
or overstate $\left( \overline{u}\right) $\ the benefit from just in time
buying 
\begin{equation*}
u_{A}\left( x_{A}\left( \theta _{A}\right) ,\theta _{A}\right) <\overline{u}%
_{A}\left( x_{A}\left( \theta _{A}\right) ,\theta _{A}\right) .
\end{equation*}%
Either an understated benefit $\underline{u}_{A}\left( x^{\ast }\left(
\theta \right) ,\theta _{A}\right) $ of the long-term buying or an
overstated benefit $\overline{u}_{A}\left( x_{A}\left( \theta _{A}\right)
,\theta _{A}\right) \ $of the short-term buying, leads to a larger transfer from 
the supplier, since 
\begin{equation*}
t_{S}\left( x,\theta \right) =u_{A}\left( x_{A}\left( \theta _{A}\right) ,\theta
_{A}\right) - u_{A}\left( x^{\ast }\left( \theta \right)
,\theta _{A}\right)  <  \overline{u}_{A}\left( x_{A}\left( \theta _{A}\right) ,\theta
_{A}\right) - \underline{u}_{A}\left( x^{\ast }\left( \theta \right) ,\theta
_{A}\right) \text{,}
\end{equation*}%
but it may also lead to an inefficient decision  since from (\ref{eff}%
): 
\begin{equation*}
\underset{x\in X}{\arg \max }\left\{ \overline{u}_{A}\left( x,\theta
_{A}\right) +u_{S}\left( x,\theta _{S}\right) \right\} \neq \ \underset{x\in X}{%
\arg \max }\left\{ u_{A}\left( x,\theta _{A}\right) +u_{S}\left( x,\theta
_{S}\right) \right\} .
\end{equation*}%
Yet the advantage of this method is that we can adapt the transfer of the
payment to the size of the gains, that is we can choose how to determine $%
\overline{u}_{A}\left( x,\theta _{A}\right) $, with the three leading
alternatives: \emph{additive boosting}%
\begin{equation*}
\overline{u}_{A}\left( x,\theta _{A}\right) =u_{A}\left( x,\theta
_{A}\right) +\alpha , \label{p1}
\end{equation*} 
\emph{multiplicative boosting}
\begin{equation*}
\overline{u}_{A}\left( x,\theta _{A}\right) =(1+ \beta) \cdot u_{A}\left( x,\theta
_{A}\right) 
\end{equation*}%
or a \emph{return on investment (ROI)} based activity fee 
\begin{equation*}
\overline{u}_{A}\left( x,\theta _{A}\right) =  u_{A}\left( x,\theta
_{A}\right) + r \cdot |(u_{A}\left( x,\theta
_{A}\right) - u_{A}\left( x^{\ast }\left( \theta \right)
,\theta _{A}\right))|
\end{equation*}
for $\alpha, \beta, r >0 $. 
To illustrate, in the additive case, the value of the supplier's VCG transfer is now 
\[
t_{S}\left( x,\theta \right) =u_{A}\left( x_{A}\left( \theta _{A}\right) ,\theta_{A}\right) - u_{A}\left( x^{\ast }\left( \theta \right) ,\theta _{A}\right) + \alpha ,
\]
\noindent so we have simply increased the transfer payment of the supplier by $\alpha$. The additive payment preserves the efficiency of the coordinated outcome because it only shifts the utility function by a constant; the multiplicative and ROI versions do not preserve efficiency but
allow us to more uniformly tie the payment to the coordination gains.

\subsection{Activity Fee: Bias Expressed in Transaction Volume}

We have expressed the benefit for the allocation in abstract terms by
utility function $u_{A}\left( x_{A,}\theta _{A}\right) ,u_{S}\left(
x_{S,}\theta _{S}\right) $, and this is a natural start as the allocation
policy can be high dimensional and we may not know how the variables
interact with the cost and transportation structure, and all the supply
chain constraints of the supplier. But for the moment, the coordination
planning is mostly about the order quantity. We can therefor add a linear
term for any deviation from jit order quantity 
\begin{equation}
c\left( x^{\ast }\left( \theta \right) ,x_{A}\left( \theta _{A}\right)
\right) =\left\{ 
\begin{array}{ccc}
\overline{c}\left( x^{\ast }\left( \theta \right) -x_{A}\left( \theta
_{A}\right) \right)  & \text{if} & x^{\ast }\left( \theta \right) \geq
x_{A}\left( \theta _{A}\right)  \\ 
\underline{c}\left( x_{A}\left( \theta _{A}\right) -x^{\ast }\left( \theta
\right) \right)  & \text{if} & x^{\ast }\left( \theta \right) <x_{A}\left(
\theta _{A}\right) 
\end{array}%
\right.   \label{eq:lp}
\end{equation}%
We can choose the linear penalities $0\leq \underline{c}\leq \overline{c}%
<\infty \,\ $to give different penalties to over and undersupply
respectively. We can also vary $c_{i,t}$ with the asin $i$ and the time $t$. 

We can now introduce the bias into the utility function, thus%
\begin{equation*}
\underline{u}_{A}\left( x^{\ast }\left( \theta \right) ,\theta _{A}\right)
=u_{A}\left( x^{\ast }\left( \theta \right) ,\theta _{A}\right) -c\left(
x^{\ast }\left( \theta \right) ,x_{A}\left( \theta _{A}\right) \right) .
\end{equation*}%
Now for constant $c$, we are collecting more than our status quo surplus.
Also we can vary $c$ over time and learn how responsive the social decision
is the level of $c$.  If we a coordinated plan now deviates from the status
quo, or 
\begin{equation*}
x^{\ast }\left( \theta \right) >x_{A}\left( \theta _{A}\right) \text{,}
\end{equation*}%
then we know the value for the supplier is at least 
\begin{equation*}
\overline{c}\left( x^{\ast }\left( \theta \right) -x_{A}\left( \theta
_{A}\right) \right) \text{,}
\end{equation*}%
and hence this is lower bound on the gains to the supplier that we could
appropriate.  

%\newpage 

\section{Dynamic Allocation}
We currently formulate the VCG allocation problem as a static, single-period problem. In reality, demand predictions, supply conditions and expectations can change over time. In this section, we extend the insights of the VCG mechanism in a rolling T-week horizon framework ($T=6$)  to dynamic environments \footnote{  We will introduce some new notations for clarity, and we ignore the activity fee in this section for  simplicity}. A dynamic version of the VCG mechanism, called the dynamic pivot mechanism, preserves the efficiency and simplicity of the static mechanism, see \cite{beva10} and \cite{beva19}.

\subsection{Baseline: Ideal JIT Policy}

The baseline alternative can be chosen as the \emph{ideal JIT policy}, which involves choosing order quantities that exactly bring the inventory position up to the target inventory position (TIP) in each period. The ‘ideal JIT policy’ serves as a theoretical benchmark; real-world implementations may deviate due to forecasting errors, capacity constraints, or operational flexibility. We consider a retailer ordering from a supplier, where the ordering policy $\{z_t\}$ maximizes the utility of a retailer, given in flow terms as $a_t : \mathbb{R}_+ \to \mathbb{R}$. At planning moment $s$, we have information about inventory, demand, supply, etc., and interpret the flow utility as the incremental utility arising from an order. It is therefore useful to condition the flow utility on the information given $s$: $\max_{z_t} \sum_{t=s}^{s+5} a_t(z_t \mid s)$.In the independent ordering system, the optimal order sequence leads to a purchase order $z_s$ for period $s$ and a set of associated forecasts $z_t = f_t$ for $t > s$. For the six-week planning horizon, we denote this sequence as: $z_s, \hat{z}_{s+1}, \ldots, \hat{z}_{s+5}$, where $\hat{z}_{s+k}$ denotes the forecast of the JIT order quantity for period $s+k$.

\subsection{Coordinated Ordering}

Coordination enables the PO plan to deviate from the ideal JIT orders by incorporating the supplier's private cost information. Let $v_t : \mathbb{R}_+ \to \mathbb{R}$ denote the supplier's flow utility. The retailer and the supplier attempt to solve: $\max_{x_t} \sum_{t=s}^{s+5} \left[ a_t(x_t \mid s) + v_t(x_t \mid s) \right].$ In the coordinated ordering system, the optimal order sequence leads to a purchase order $x_s$ for period $s$, agreed future plans $x_{s+k}$ for $k = 1, \ldots, 5$, and a set of associated forecasts $z_t = f_t$ for $t > s + 5$. We denote this consensus plan as: $ x^{*}_s, x^{*}_{s+1}, \ldots, x^{*}_{s+5}$. At subsequent period $s+1$, we allow re-coordinating the PO quantities using updated information about supply and demand.

\subsection{Commitment Structure and Transfer Payment under Rolling Horizon}
We consider the case where in week $t$, we do not issue purchase orders beyond current period and these future PO quantities are not explicit commitments. After reaching the consensus, 
${x}_{t}^{\ast },... ,{x}_{t+5}^{\ast }$, we issue the first period PO, ${x}_{t}^{\ast }$. Then  we compute what the resulting future JIT policy quantities would be as a result of placing the initial order, which we denote by $(\overline{z}_{t+1},...,\overline{z}_{t+5})$. That is, these are the POs that would maximize a retailer's utility over the horizon conditioned on using the first period PO consensus  $x_{t}^{\ast }$. We compare this utility to the utility of the ideal JIT policy (with no constraint on ${x}_{t}^{\ast }$) to determine the transfer, which we term the Cost Benefit Transfer (CBT). Thus the CBT payment from the supplier to the retailer at period $t=s$ is: 
\begin{equation}
CBT^{(1)}(s) = \left[ a_s(z_s \mid s) +\sum_{t=s+1}^{s+5} a_t(\hat{z}_t \mid s) \right] - \left[ a_s(x^{*}_s \mid s) + \sum_{t=s+1}^{s+5} a_t(\bar{z}_t \mid s) \right]\\\\
%\doteq u_A(z_s, \hat{z}_{s+1}, \ldots, \hat{z}_{s+5} \mid s) - u_A(x^*_s, \bar{z}_{s+1}, \ldots, \bar{z}_{s+5} \mid s).
 \geq 0 \end{equation}
where the inequality follows from the definition of the ideal JIT policy.
Thus in every period in which we see a deviation from the JIT policy, we have a positive CBT payment from the supplier to the retailer.
The payment can be thought of as the immediate cost of the week $t$ deviation plus the certainty equivalent cost of the deviation on future periods. \footnote{Note that the current formula implicitly assumes a 100\% supplier fill rate. Relaxing this assumption allows CBT to be either positive or negative; however, we retain the current one-way formulation as designing a manipulation-resistant counterfactual baseline for two-way CBT remains an open challenge. One can interpret this idealization of JIT cost as a form of utility “squashing”.}

If we have multi-period commitments, then we would have to price them based on the expected payments that we would have to make to the supplier due to changes in the private values of a retailer and the associated adjustments in POs. In this mental model, the consensus PO plan over a 6-week horizon from a given plan week is treated as a commitment in future weeks. Specifically, as we roll forward a week, prior commitments define the updated baseline plan for the first five weeks of the horizon and a sixth week of a retailer-chosen PO’s is appended to complete the baseline plan. A new consensus is then run and CBT payments are computed based on the difference between the cost of the consensus and baseline plans. Below we show the formula details. 

Under full 6-week commitment, all six weeks of the consensus plan are binding. Suppose starting from period $t=s$, if there are agreements about the next 6 weeks then the CBT payment for the vendor is:
\begin{equation}
   CBT^{(6)}(s) =  \left[a_s(z_s \mid s) +\sum_{t=s+1}^{s+5} a_t(\hat{z}_t \mid s) \right]  - \sum_{t=s}^{s+5} a_t(x^{*}_t \mid s).
\end{equation}

In the subsequent period $s+1$, the prior consensus quantities $x^{*}_{s+1},..., x^{*}_{s+5}$ become binding commitments (the plan of record). The updated baseline thus takes the form: $x^{*}_{s+1}, x^{*}_{s+2}, \ldots, x^{*}_{s+5}, \hat{z}{s+6}$, where only the final term $\hat{z}_{s+6}$ is freely optimized to maximize a retailer's utility. Denote solution sequence of the new independent plan is $z_{s+1}, \hat{z}_{s+2}, \ldots, \hat{z}_{s+6}$. 

In the subsequent period $s+1$, a retailer and supplier again maximize the joint utility value. Denote the solution to the new coordinated plan as $x^{*}_{s+1},..., x^{*}_{s+6}$.

The agreement can be updated and a monetary payment is made:%
\begin{equation}
CBT^{6}\left( s+1\right) =\left[a_{s+1}(z_{s+1} \mid s) +\sum_{t=s+2}^{s+6} a_t(\hat{z}_t \mid s) \right]  - \sum_{t=s+1}^{s+6}a_{t}\left( x^{*}_{t}\left\vert s+1\right.
\right)  .
\label{ov}
\end{equation}

For the week after, we repeat the above algorithm to update both the baseline plan, and the coordinated plan, as well as CBT.

\section{VCG Implementation Alternatives}
Besides the direct VCG mechanism, we have alternative
indirect means to obtain the same outcome. The desirability of the mechanism will
depend on the complexity and dimensionality of the problem, and the
amount of information that the suppliers have. Here are two
alternative implementations.

\subsection{Menu of alternatives}

Notice that prices offered to supplier do not depend on the information of the
supplier but on the information of a retailer, see (\ref{ts}) $t_{S}\left( x^{\ast }\left( \theta \right) ,\theta \right) = u_{A}\left( x_{A}\left(
\theta _{A}\right) ,\theta _{A}\right) - u_{A}\left(
x^{\ast }\left( \theta \right) ,\theta _{A}\right) + \alpha $.
We therefore can offer a menu of plans and associated prices: 
\begin{equation}
t_{S}\left( x,\theta \right) = u_{A}\left(
x_{A}\left( \theta _{A}\right) ,\theta _{A}\right) - u_{A}\left( x,\theta _{A}\right) + \alpha.  \label{vm}
\end{equation}%
for all $x\in X$, and not only $x^{\ast }\left( \theta \right) $. This would
be the menu that the supplier is being offered and by choosing the item on the
menu that they likes best, they would maximize social surplus.

\noindent \textbf{Theorem 1 (Menu VCG). } {\em The modified VCG\ mechanism can be
attained by VCG\ menu (as in (\ref{vm})) where the agents respond in dominant
strategies.}

Finally, many times we may not know what the feasible alternatives for the
suppliers are. In this case, we might just elicit feasible options from them,
call them $X_{f}$, and then we can price just the feasible options. This
would make the evaluation simpler and shorter and would yield again the
optimal outcome.

\noindent \textbf{Theorem 2 (Supplier Initiated VCG). } {\em The modified VCG\
mechanism can be attained by VCG\ menu (as in (\ref{vm}) where a retailer offers
transfers and prices for all $x\in X_{f}$ and the agents respond in dominant
strategies.}

\subsection{Iteration using CPP}
This is the version described above.
The VCG\ mechanism asks for a complete value profile. Using CPP, a supplier’s “value profile” is the CPP agent it submits to the platform, and CPP is the solution engine that computes the optimal consensus plan. 
Submitting truthful agents (i.e., agents that truthfully optimize their utility in response to CPP queries) is a dominant bidding strategy.

\noindent \textbf{Theorem 3 (Iterative VCG). } {\em The modified VCG\ mechanism
can be attained by a price search of CPP where the agents all respond in
dominant strategies.}

An advantage of using CPP is that agents do not have to explicitly reveal their complete utility functions, but rather can provide a “black box” software agent/API that solves their agent utility maximization problem in response to CPP coordinator queries about their preferred plans. This provides a high degree of information privacy. The implementation can be straightforward: a supplier's agent iterates with a retailer to determine a coordinated plan through CPP, receiving a single "take-it-or-leave-it" offer price that bundles both the VCG payment and the activity fee.

\section{\textbf{Toy Example of Retailer and Supplier Utility and CPP-VCG Outcomes}}
We next propose one option for designing a retailer and supplier agents. To simplify, we assume planning takes place over a single period (e.g., a quarter) and the vendor supplies a single product (ASIN). Suppose the a retailer inbound nodes are indexed by $i =1,\dots,I. $. Then the supply plan reduces to $x=[x_{i}]$ where $x_{i}$ is the quantity of product the vendor supplies to a retailer at each inbound node $i$ during the planning horizon. (The formulation can easily be extended to multiple product and multiple time periods.)

We propose representing private utilities as an optimal transport problem. Optimal transport is a good representation of utility if costs are linear, since it allows us to enumerate the various sources of supply and sources of demand, with arc costs representing the variable cost of satisfying a given demand source from a given supply source. Bounds on supply can be added to capture the limited capacity of a source, and bounds on demand can be added to capture the desire to have sufficient supply to meet demand. Indeed, this specification may be a good starting point for a pilot.

In the case of a retailer, the source nodes are the a retailer inbound nodes and the destination nodes are the demand regions where inventory is stored and demand is fulfilled. In the case of a supplier, we specify a number of potential source nodes (private information), which are production or storage locations of the supplier and the destination nodes are the a retailer inbound nodes. The math models are detailed below.

\subsection{Retailer Utility}
Suppose the retailer's outbound nodes (delivery stations/customer demand regions) are indexed by $j=1,... J$. These could be as granular as a delivery station jurisdiction or as course as a North American region. The demand for the product at outbound node $j$ is denoted as $d_j$. Let Let $d=[d_{j}]$ and let $\pi_A(d)$ denote the a retailer’s gross profit (profit excluding transportation/fulfillment costs) of satisfying demand. a retailer’s private variables are it’s transportation plan, $v^A=[v^A_{ij} ]$, where $v^A_{ij} $ is the number of units of product delivered from inbound node $i$ to outbound node $j$. There is a per-unit transport cost $c_{ij}$ associated with supplying units from $i$ to $j$, which includes are variable cost on the path from $i$ to $j$ (e.g., labor processing costs, transport costs, opportunity costs of capacity, etc.)

Given a supply plan $x$, a retailer utility for $x$ is then determined by minimizing the transportation cost to satisfy demand:

\begin{eqnarray}
 u_A(x, \theta_A) =\pi_A(d) ~~ -&  \min_{v^A} \sum_{i=1}^I \sum_{j=1}^J c^A_{ij} v^A_{ij} \\  & s.t.\sum_{i=1}^I v^A_{ij} \geq d_j, ~~ j=1,\ldots, J\\  &  \sum_{j=1}^J v_{ij}^A \leq x_i, ~~ i=1,\ldots,I \\ & v^A_{ij}\geq 0, ~~ \forall i,j 
\end{eqnarray}%

where $\theta_{A}$ denotes a retailer’s private information, which in this case is the demand and transportation cost information. This information is not known to the supplier. 

\subsection{Supplier utility}

The supplier faces a problem that is symmetric to a retailer’s. Specifically, the supplier has multiple source nodes (production facilities or warehouses) indexed by $k=1,.. K$, which can be used to supply a retailer’s inbound nodes $i=1,.. I$. Each supply node $k$ has a capacity constraint, denoted $s^S_k$, and the supplier’s per unit cost to supply inbound node $i$ from source node $k$ is denoted $c_{ki}^S$. The supplier’s private variables are its transportation plan $v^S=[v^S_{ki} ]$, which is chosen to minimize the cost of fulfilling supply plan $x$. Hence, the vendor’s utility is

\begin{eqnarray}
u_S(x, \theta_S) = \pi_S(x) ~~ - & \min_{v^S} \sum_{i=k}^K \sum_{i=1}^I c^S_{ki}  v^S_{ki} \\ & s.t. \sum_{k=1}^K v^S_{ki} \geq x_i, ~~ i=1,\ldots, I\\  &  \sum_{i=1}^I v_{ki}^A \leq s_k, ~~ k=1,\ldots,K \\ & v^S_{ki}\geq 0, ~~ \forall k,i 
\end{eqnarray}%

where $\pi_S(x)$ is the vendors gross profit (profit excluding supply chain costs) from supplying $x$, and $\theta_{S}$ denotes the supplier’s private information, which in this case is the supplier’s source node capacities and transportation costs. This information is not known to the retailer.

\subsection{Numerical example}
We next give a toy numerical example with two inbound nodes, two demand regions and two supplier source nodes.

\begin{table}[h]
    \centering
    \begin{tabular}{|r|r|r|r|r|r|}
    \hline
        Cost parameter & Demand Node 1  & Demand Node 2   \\  \hline
        Inbound Node 1 & \$1 & \$5  \\ \hline
        Inbound Node 2 & \$2 & \$3   \\ \hline
        Gross profit/unit & \$20 & \$20   \\ \hline
        Demand (units) & 40 & 60       \\ \hline
    \end{tabular}
    \caption{Retailer’s private information}
\end{table}

\begin{table}[h]
    \centering
    \begin{tabular}{|r|r|r|r|r|r|}
    \hline
        Cost parameter & Inbound node 1  & Inbound node 2 & Capacity  \\  \hline
        Source node 1 & \$10 & \$5 & 100  \\ \hline
        Source node 2 & \$1 & \$2 & 10 \\ \hline
        Gross profit/unit & \$20 & \$20 &      \\ \hline
    \end{tabular}
    \caption{Supplier’s private information}
\end{table}

\subsection{Retailer JIT Ordering}
Under the cost parameters, a retailer’s lowest cost is to to use Inbound node 1 to fulfill demand at Demand node 1 and to use Inbound node 2 to fulfill Demand node 2. Hence, a retailer would prefer the supply plan $x_1^0 = 40,  x_2^0 = 60$ and the corresponding optimal transportation plan  $v^A_{11} = 40, v^A_{1,2} = 0, v^A_{21} = 0, v^A_{22} = 60$. The total cost for a retailer is $(\$1\times40+\$3\times60)=\$220$ and its gross profit is $\$20\times(40+60)=\$2,000$, so it’s utility is $\$1,780$. We refer to this as the a retailer JIT plan.\\

To satisfy the order $x_1^0 = 40,  x_2^0 = 60$, the supplier would choose its optimal transportation plan subject to its capacity constraints, which is  $v^S_{11} = 30, v^S_{1,2} = 60, v^S_{21} = 10, v^S_{22} = 0$ . The total cost for the supplier is $\$610$,  and its gross profit is $\$20\times(40+60)=\$2,000$, so its utility is $\$1,390$.  \\

Note in this situation, the total supply chain cost (retailer + supplier) equals $\$220+\$610=\$830$. 

\subsection{Value of Coordination (First-Best)}
Suppose there exists a social planner that is aware of all private information and can solve for the minimum supply chain cost subject to the customer demand constraints and the supplier capacity constraints. Specifically,   

\begin{equation*}
\max_x \{ u_A(x,\theta_A) + u_S(x, \theta_S) \}
\end{equation*}%

The optimal social first-best solution is: $x_1^* = 10,  x_2^* = 90$, with corresponding private variable choices of $v^A_{11} = 10, v^A_{1,2} = 0, v^A_{21} = 30, v^A_{22} = 60$ for a retailer and  $v^S_{11} = 0, v^S_{1,2} = 90, v^S_{21} = 10, v^S_{22} = 0$ for the vendor. The gross profit of a retailer and the vendor are the same as before. So a retailer’s utility becomes $ \$2,000 - \$250=\$1,750$, and the vendor’s utility becomes $ \$2,000 - \$460=\$1,540$.  

However, the total supply chain cost of a retailer and the vendor is now $\$710$, which is $\$120$ lower than the JIT cost of $\$830$ ($14.5\%$ lower total supply chain cost).  This means the gains from coordination are $g(\theta)=\$120$.

Note that now, a retailer incurs a total transportation cost of $\$250$, which is $\$30$ greater than a retailer’s transportation cost of $\$220$ under JIT. The vendor incurs a total transportation cost of $\$460$, which is $\$150$ lower than it’s JIT cost of $\$610$ under JIT ($25\%$ lower). 

\subsection{VCG Payments}

The supplier’s VCG payments are then $\$30+\alpha$, which is equal to a retailer’s increased cost ($\$30$) plus the fixed activity fee $\alpha$. The CPP-VCG mechanism would compute the optimal plan, $x^*$, and then offer it to the supplier at a price of $\$30+\alpha$. Since the vendor’s cost is $\$150$ lower under the plan $x^*$, the supplier would take the offer if $\$150 > \$30 + \alpha$ or equivalently $\alpha < \$120$. For example, suppose a retailer set the activity fee at $\alpha = \$50$, then the supplier would participate since is has a surplus of $\$150 -( \$30 +\$50) = \$70$, and a retailer would generate a surplus of $\$50$. The combined surplus is $\$120$, which is total gains from coordination. 

\subsection{Menu  of alternatives VCG}
Suppose that vendor is willing to share us information that their transportation cost sending to Inbound node 2 is on average cheaper than sending to inbound node 1, and thus asks if we can increase the inbound supply mix to inbound 2. Received this information, and under the same activity fee $\alpha =\$50$, a retailer may come back with menu-of-contracts set of options like: 
\begin{enumerate}
 \item Supply plan 1 : $[x_1 = 30, x_2 = 70]$, with fee $\$60$; 
 \item Supply plan 2: $[x_1 = 20, x_2 = 80]$, with fee $\$70$;
 \item Supply plan 3: $[x_1 = 10, x_2 = 90]$, with fee $\$80$;
 \item Supply plan 4: $[x_1 = 0, x_2 = 100]$, with fee $\$90$.
\end{enumerate}

, where for a plan $x'$ in the menu, we use $t_s(x, \theta) = u_A([x_1=40, x_2=60], \theta_A) - u_A(x', \theta_A)+\alpha $ to compute the fees above.  

The vendor can simply respond in dominant strategy by choosing the option that yield the highest utility for the vendor.  Specially, the vendor’s optimal private plan and corresponding utilities associated with four plans are: 
\begin{enumerate}
 \item Under supply plan 1, vendor would choose $[v^{S'}_{11}=20, v^{S'}_{12}=70,v^{S'}_{21}=10,v^{S'}_{22}=0]$, and obtain utility  $\$2000 - \$560 - \$50 = \$1390 $; 
 \item Under supply plan 2,   vendor would choose  $[v^{S'}_{11}=10, v^{S'}_{12}=80,v^{S'}_{21}=10,v^{S'}_{22}=0]$ , and obtain utility $\$2000 - \$510 - \$50 = \$1440 $;
 \item Under supply plan 3,   vendor would choose  $[v^{S'}_{11}=0, v^{S'}_{12}=90,v^{S'}_{21}=10,v^{S'}_{22}=0]$ , and obtain utility $\$2000 - \$460 - \$50 = \$1490  $;
 \item Under supply plan 2,   vendor would choose  $[v^{S'}_{11}=0, v^{S'}_{12}=90,v^{S'}_{21}=0,v^{S'}_{22}=10]$ , and obtain utility $\$2000 - \$470 - \$50 = \$1480 $.
\end{enumerate}

The vendor would choose the supply plan 3 $[x_1 = 10, x_2 = 90]$ that comes with fee $\$80$, which yields the highest vendor’s utility among all the options (and higher than the vendor’s under under the JIT status quo plan). This example shows that we can use menu-based approach to achieve the socially optimal first-best solution as well, particularly when effective information sharing facilitate the generation of promising options.

\section {Concluding Remarks}
This paper presents the CPP-VCG theoretical framework. This novel approach marks a significant departure from conventional siloed optimization methods, enabling retailers and their suppliers to reduce total cost to serve and to improve supply chain efficiency.

The CPP-VCG framework is, at its core, a general-purpose tool for achieving consistent outcomes when each party acts on information unavailable to the other. The supply chain application we describe here is a natural first use case, but the underlying logic extends well beyond it. Vendor negotiations, third-party logistics partnerships, and multiparty logistics planning all involve the same fundamental structure: interdependent decisions, private costs, and benefits that can be unlocked only through carefully designed incentive-compatible mechanisms.

Several open research questions remain. How should the mechanism handle the situation in which the suppliers face supply shortages and cannot deliver the agreed-upon quantities? Can mutual commitment structures — where both parties share obligations — improve outcomes when forecasts are highly uncertain? These are questions that sit at the intersection of economic theory and large-scale systems engineering. Addressing these questions will not only refine the practical implementation of the CPP-VCG framework but also deepen our understanding of how incentive-compatible mechanisms can unlock efficiency gains in real-world supply chains and multi-party logistics networks where information asymmetry and interdependent decisions prevail.

\newpage 

\bibliographystyle{plainnat}
\bibliography{general} 
%\printbibliography

\newpage

\end{document}